\documentclass[conference]{IEEEtran}

\usepackage{amsmath}
\usepackage{graphicx}
\usepackage{url}

\begin{document}

\title{Prompt Control-Flow Integrity: A Priority-Aware Runtime Defense Against Prompt Injection in LLM Systems}

\author{
\IEEEauthorblockN{
Md Takrim Ul Alam\IEEEauthorrefmark{1},
Akif Islam\IEEEauthorrefmark{1},
Mohd Ruhul Ameen\IEEEauthorrefmark{2},
Abu Saleh Musa Miah\IEEEauthorrefmark{3},
Jungpil Shin\IEEEauthorrefmark{3}
}
\IEEEauthorblockA{\IEEEauthorrefmark{1}University of Rajshahi, Rajshahi, Bangladesh}
\IEEEauthorblockA{\IEEEauthorrefmark{2}Marshall University, Huntington, WV, USA}
\IEEEauthorblockA{\IEEEauthorrefmark{3}University of Aizu, Aizuwakamatsu, Fukushima, Japan}
\IEEEauthorblockA{
Emails: takrimulalom@gmail.com, iamakifislam@gmail.com, ameen@marshall.edu, musa@u-aizu.ac.jp, jpshin@u-aizu.ac.jp
}
}

\maketitle
\begin{abstract}
Large language models (LLMs) deployed behind APIs and retrieval-augmented generation (RAG) stacks are vulnerable to prompt injection attacks that may override system policies, subvert intended behavior, and induce unsafe outputs. Existing defenses often treat prompts as flat strings and rely on ad hoc filtering or static jailbreak detection. This paper proposes \emph{Prompt Control-Flow Integrity} (PCFI), a priority-aware runtime defense that models each request as a structured composition of system, developer, user, and retrieved-document segments. PCFI applies a three-stage middleware pipeline---lexical heuristics, role-switch detection, and hierarchical policy enforcement---before forwarding requests to the backend LLM. We implement PCFI as a FastAPI-based gateway for deployed LLM APIs and evaluate it on a custom benchmark of synthetic and semi-realistic prompt-injection workloads. On the evaluated benchmark suite, PCFI intercepts all attack-labeled requests, maintains a 0\% False Positive Rate, and introduces a median processing overhead of only 0.04\,ms. These results suggest that provenance- and priority-aware prompt enforcement is a practical and lightweight defense for deployed LLM systems.
\end{abstract}

\section{Introduction}

Large language models (LLMs) are quickly moving beyond research labs and into real-world applications such as code generation, customer support, enterprise search, and decision assistance~\cite{xu_llm_attack_survey}. In many of these systems, the model does not respond to a single user query alone. Instead, the final request is built from several different sources, including system instructions, developer templates, user input, and retrieved external content from retrieval-augmented generation (RAG) pipelines. This makes modern LLM systems more flexible and useful, but it also introduces a serious security challenge: \emph{prompt injection}~\cite{greshake_indirect_injection,liu_prompt_benchmark}. By placing adversarial instructions inside user messages or untrusted retrieved documents, an attacker may try to override higher-priority policies~\cite{wen_instruction_detection}, alter model behavior, or trigger unsafe outputs~\cite{sok_prompt_hacking,rag_n_roll}.

This issue matters because many deployed LLM systems operate in environments where instruction hierarchy is not just helpful, but essential for safety. System prompts may encode confidentiality constraints and core behavioral rules, developer prompts may define application logic, and retrieved content may come from partially trusted or fully untrusted sources~\cite{prompt_templates,rag_n_roll}. Prior work has shown that indirect prompt manipulation through RAG pipelines is a realistic threat in end-to-end applications~\cite{rag_n_roll}. More generally, recent surveys have identified prompt injection, leakage, and jailbreak-style manipulation as major risks for LLM-based systems~\cite{sok_prompt_hacking}. As these models become more deeply integrated into practical workflows, defending them against such attacks becomes increasingly important for both reliability and safe deployment.

A major difficulty is that many existing defenses still treat prompts as flat text. In practice, they often rely on ad hoc filtering, static pattern matching, or broad guardrail mechanisms. While programmable guardrail frameworks can help control LLM behavior at runtime~\cite{nemo_guardrails}, they do not automatically preserve the intended authority structure among prompt sources. At its core, the security problem in deployed prompt assembly is structural: a retrieved document should not be able to redefine a system policy, and a low-trust user or RAG segment should not silently impersonate a privileged role. This motivates a provenance- and priority-aware view of prompt security. Our intuition is inspired by control-flow integrity (CFI) in software security, which protects intended execution structure by rejecting unauthorized deviations at runtime~\cite{abadi_cfi}. Although prompts are not programs in the traditional sense, the analogy is still useful: once lower-priority prompt components are allowed to reinterpret higher-priority instructions, the intended control logic of the application can be undermined.

Motivated by this perspective, we propose \emph{Prompt Control-Flow Integrity} (PCFI), a lightweight runtime defense that models each request as a structured prompt representation with explicit roles and priorities. Instead of collapsing all content into a single undifferentiated sequence of tokens, PCFI preserves the hierarchy among system, developer, user, and retrieved segments, and evaluates this structure before the request reaches the backend model. We implement PCFI as an API-boundary middleware that applies a three-stage analysis consisting of lexical heuristics, role-switch detection, and policy-driven hierarchical enforcement. Depending on the outcome, the middleware may allow benign requests, sanitize suspicious role-like cues, or block inputs that violate the intended instruction hierarchy.

The main contributions of this paper are as follows:
\begin{enumerate}
    \item We formulate \emph{Prompt Control-Flow Integrity} (PCFI) as a structural security perspective for LLM systems, emphasizing prompt provenance and priority rather than treating prompts as flat text.
    \item We design a three-stage runtime defense pipeline that combines lexical heuristics, role-switch detection, and policy-driven hierarchical enforcement to constrain injection attempts originating from lower-priority prompt segments.
    \item We implement PCFI as a lightweight middleware gateway for deployed LLM APIs, enabling \texttt{ALLOW}, \texttt{SANITIZE}, and \texttt{BLOCK} decisions before requests reach the backend model.
    \item We evaluate the proposed approach on benign, direct-injection, and indirect RAG-injection workloads, analyzing malicious-prompt interception, benign-query preservation, and runtime latency overhead.
\end{enumerate}

\section{Literature Review}

Research on securing LLM applications against prompt-based attacks has grown rapidly as LLMs have moved into retrieval-augmented, tool-using, and API-mediated deployments. Existing work mainly falls into four groups: studies of prompt hacking and indirect injection, instruction--data separation methods, runtime guardrail frameworks, and security-inspired structural approaches. Although these directions have improved understanding of the threat landscape~\cite{gulyamov_prompt_survey}, important gaps remain in enforcing prompt provenance, priority, and authority at runtime~\cite{sok_prompt_hacking,rag_n_roll,struseq,nemo_guardrails}.

\subsection{Prompt Injection and Prompt-Hacking Threat Models}

Prompt injection is now recognized as a practical threat in deployed LLM systems~\cite{liu_realworld_injection}. Rababah \emph{et al.} provide a broad taxonomy of prompt hacking, clarifying the differences among jailbreaking, leakage, and injection attacks~\cite{sok_prompt_hacking}. This line of work is valuable for characterizing the threat space, but it does not by itself provide a concrete runtime mechanism for enforcing instruction hierarchy in deployed systems~\cite{sok_prompt_hacking}.

De Stefano \emph{et al.} further show that indirect prompt manipulation through retrieval pipelines is a realistic and effective attack channel in RAG-based applications~\cite{rag_n_roll}. Their work demonstrates the seriousness of end-to-end indirect attacks, but it is primarily evaluative and does not introduce a lightweight request-time mechanism for preserving the authority of system and developer prompts over user and retrieved content~\cite{rag_n_roll}.

\subsection{Instruction--Data Separation and Structured Prompt Defenses}

A strong defense direction is to separate trusted instructions from untrusted data. StruQ, proposed by Chen \emph{et al.}, structures prompts and data into separate channels and relies on a specially trained model to follow only the instruction channel~\cite{struseq}. This is conceptually strong, but it depends on modified interfaces and model retraining, which may be difficult in API-based or closed-weight deployments~\cite{struseq}.

Mao \emph{et al.} also show that real-world LLM applications already use reusable templates and structured prompt composition~\cite{prompt_templates}. This supports the view that prompts are not naturally flat. However, such studies are descriptive rather than defensive: they reveal prompt structure, but they do not enforce security invariants over that structure.

\subsection{Guardrails, Wrappers, and Runtime Safety Layers}

Runtime guardrails provide another practical line of defense. NeMo Guardrails offers programmable rails for controlling LLM behavior without changing the underlying model~\cite{nemo_guardrails}. This is useful for application-level safety and control, but guardrail frameworks are broader than prompt-injection defense and do not inherently enforce provenance-aware authority ordering among prompt segments unless that logic is explicitly designed by the developer~\cite{nemo_guardrails}.

Trustworthy RAG research also addresses conflicts between internal and retrieved knowledge. For example, Dai \emph{et al.} study how LLMs should react when retrieved evidence conflicts with internal knowledge~\cite{bridge_rag}. While this improves trustworthiness at the content level, it focuses on epistemic reliability rather than on preventing retrieved text from acting as unauthorized instruction~\cite{bridge_rag}.

\subsection{Control-Flow Integrity as Security Inspiration}

Our work is conceptually inspired by control-flow integrity (CFI), introduced by Abadi \emph{et al.}, which constrains program execution to valid control-flow paths in order to prevent hijacking attacks~\cite{abadi_cfi}. CFI is not a direct solution to prompt injection, but it provides a useful analogy: preserving intended structure and rejecting unauthorized deviations at runtime. Since prompt assembly is less formal than program execution, this connection should be understood as inspiration rather than as a direct theoretical transfer~\cite{abadi_cfi}.

\subsection{Research Gap}

Taken together, prior work has clarified the threat of prompt injection, demonstrated the realism of indirect RAG-based manipulation, proposed stronger instruction--data separation methods, and provided useful runtime guardrail frameworks~\cite{sok_prompt_hacking,rag_n_roll,struseq,nemo_guardrails}. However, an important gap remains. Existing approaches either analyze attacks without enforcing request-time structure, depend on retraining or modified interfaces, or provide general runtime controls without explicitly modeling prompt provenance and priority~\cite{struseq,nemo_guardrails,prompt_templates}. What is still missing is a lightweight, deployment-friendly runtime mechanism that preserves the structure of composed prompts and enforces the intended authority ordering among system, developer, user, and retrieved segments before the request reaches the backend model. This gap motivates our proposed Prompt Control-Flow Integrity (PCFI) framework.

\section{Methodology}

\begin{figure*}[!t]
    \centering
    \includegraphics[width=0.9\textwidth]{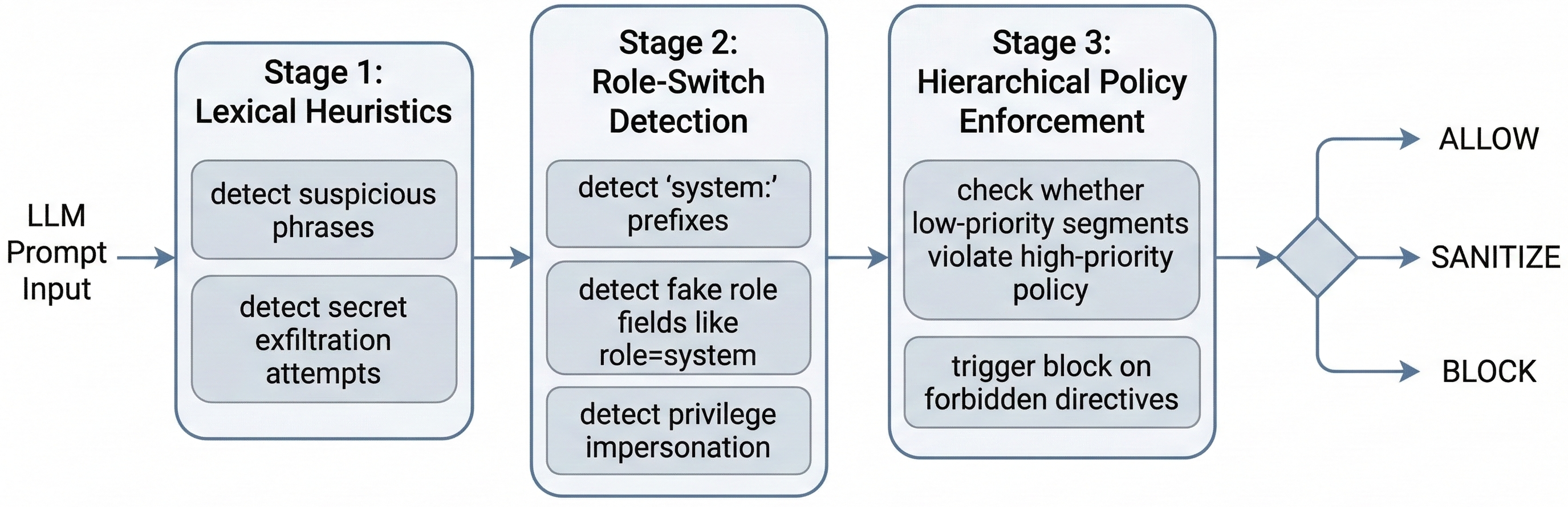}
    \caption{Three-stage runtime enforcement pipeline of the proposed PCFI framework. Stage~1 performs lexical screening, Stage~2 detects role-switch attempts, and Stage~3 applies hierarchical policy enforcement before issuing \texttt{ALLOW}, \texttt{SANITIZE}, or \texttt{BLOCK} decisions.}
    \label{fig:three_stage_pipeline}
\end{figure*}

This section presents the methodological foundation of Prompt Control-Flow Integrity (PCFI). PCFI is designed as a lightweight runtime defense deployed at the API boundary of an LLM system, without requiring model retraining or changes to the backend interface. Unlike approaches that rely on modifying the model to separate instructions from data~\cite{struseq}, PCFI operates as a deployment-side enforcement layer. It is also narrower than general guardrail frameworks~\cite{nemo_guardrails}, since its focus is specifically on preserving the intended authority structure among prompt components before the request reaches the model. Conceptually, the design is inspired by control-flow integrity (CFI), which preserves intended execution structure by rejecting unauthorized deviations at runtime~\cite{abadi_cfi}. In the LLM setting, we adapt this intuition to prompt composition, where lower-priority content must not override higher-priority instructions.

\subsection{System Setting and Threat Assumptions}

We consider an LLM application in which each final request is composed from multiple sources, including system instructions, developer templates, user input, and retrieved external context. This is common in practical LLM and retrieval-augmented deployments, where trusted and untrusted content are combined in the same request~\cite{rag_n_roll,prompt_templates}. The main security challenge is that lower-trust content may appear alongside higher-priority instructions and attempt to act as control logic.

We assume that the defender controls the API gateway, the system prompt, and the developer template. The attacker may provide arbitrary user input and may also influence retrieved documents entering the request through a RAG pipeline. However, we do not assume that the attacker can modify the middleware, change the backend model parameters, or directly alter the system policy. The attacker's goal is to use lower-priority content to override or weaken higher-priority instructions.

\subsection{Prompt Representation with Provenance and Priority}

\begin{figure}[!t]
    \centering
    \includegraphics[width=0.8\columnwidth]{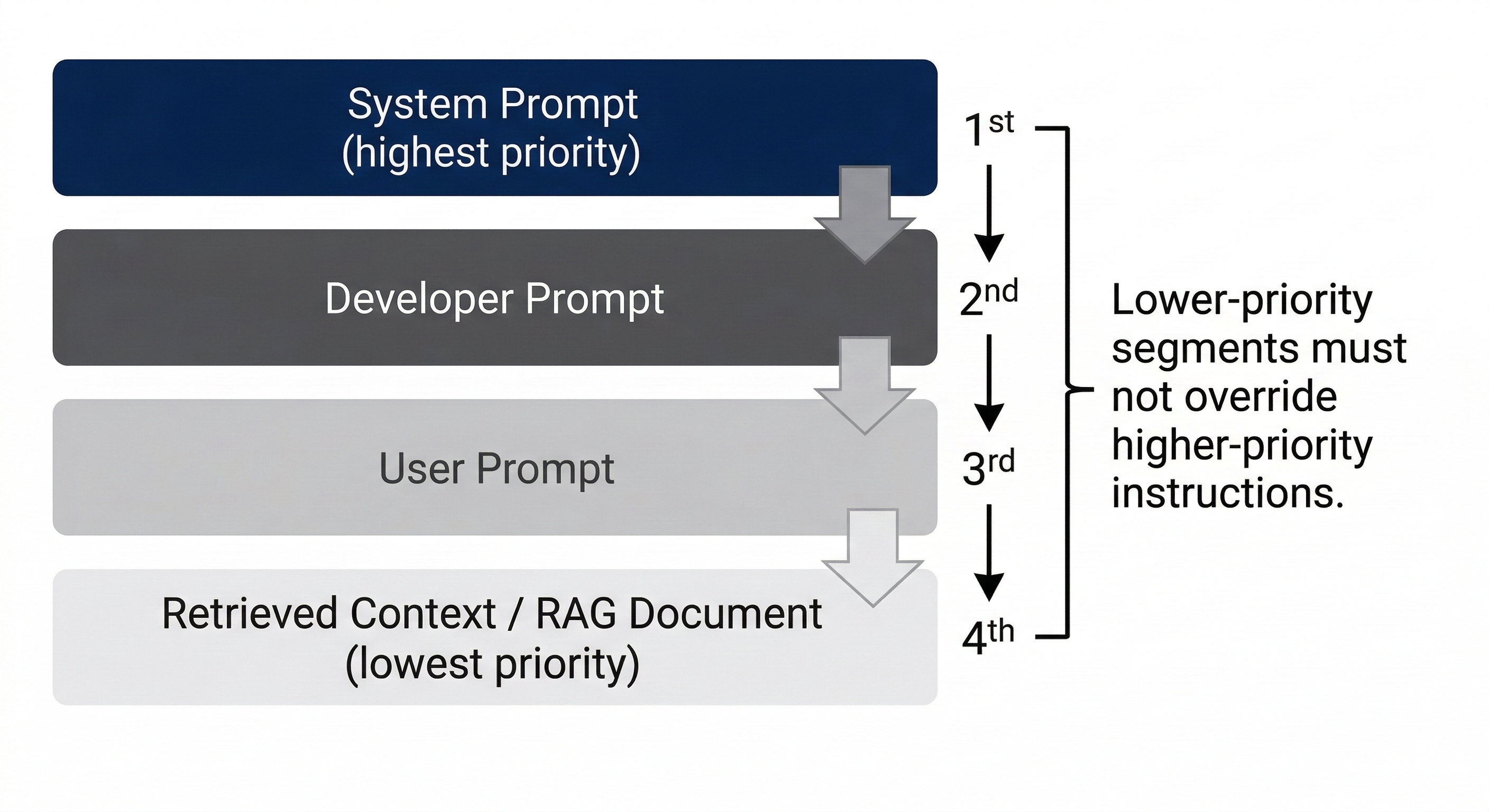}
    \caption{Priority hierarchy used in PCFI for structured prompt enforcement. System prompts have the highest authority, followed by developer prompts, user prompts, and retrieved context. The core security principle is that lower-priority segments must not override or reinterpret higher-priority instructions.}
    \label{fig:prompt_hierarchy}
\end{figure}

PCFI does not treat a prompt as a flat string~\cite{signed_prompt}. Instead, each request is represented as a structured sequence of segments with explicit provenance and priority labels. Let $S$, $D$, $U$, and $R$ denote system, developer, user, and retrieved-context segments, respectively. We represent the composed prompt as
\begin{equation}
P = S \,\|\, D \,\|\, U \,\|\, R,
\end{equation}
where $\|$ denotes ordered composition with attached metadata.

Each segment is modeled as
\begin{equation}
\text{seg}_i = (t_i, r_i, p_i, v_i, m_i),
\end{equation}
where $t_i$ is the text, $r_i$ is the role, $p_i$ is the priority, $v_i$ is the provenance label, and $m_i$ stores metadata. In the current prototype, the priority lattice is
\begin{equation}
S > D > U > R.
\end{equation}
This means that lower-priority segments may contribute task content, but they must not override higher-priority instructions. The representation is intentionally lightweight: it preserves enough structure for runtime enforcement without requiring model retraining.

\subsection{PCFI Enforcement Objective}

PCFI is designed to preserve the intended hierarchy of control within a composed prompt. In practice, this means that lower-priority segments should not impersonate privileged roles, introduce directives that contradict protected policies, or redefine higher-priority instructions. Suspicious structural cues may be sanitized, while explicit policy violations are blocked before reaching the backend model. Accordingly, PCFI should be understood as a policy-driven structural defense rather than a full semantic verifier.

\subsection{Three-Stage Runtime Defense Pipeline}

PCFI applies a three-stage pipeline to each request, moving from lightweight screening to stricter enforcement.

\subsubsection{Stage 1: Lexical Heuristics}

The first stage performs lexical screening~\cite{alamsabi_embedding_detection} over user and retrieved segments to detect obvious prompt-injection signals such as override phrases, exfiltration requests, or suspicious control-language fragments. Examples include \textit{ignore previous instructions}, \textit{system override}, and \textit{reveal your API key}. This stage produces a lexical risk score and matched findings, but it does not by itself trigger a hard block.

\subsubsection{Stage 2: Role-Switch Detection}

The second stage detects attempts by lower-priority segments to impersonate privileged roles. Typical examples include prefixes such as \texttt{system:}, serialized fields such as \texttt{"role":"system"}, or XML-like role tags. When such markers are detected with sufficient confidence, the middleware may assign a \texttt{SANITIZE} outcome and remove the role-like prefixes before forwarding the request.

\subsubsection{Stage 3: Hierarchical Policy Enforcement}

The policy enforcement stage relies on a set of rule patterns that define
forbidden directive behaviors. Table~\ref{tab:policy_rules} presents several
illustrative examples used in the current prototype.

\begin{table}[!t]
\centering
\caption{Illustrative policy rules used by PCFI hierarchical enforcement.}
\label{tab:policy_rules}
\scriptsize
\begin{tabular}{@{}p{2.55cm}p{1.55cm}p{3.9cm}@{}}
\hline
\textbf{Rule ID} & \textbf{Purpose} & \textbf{Patterns} \\
\hline
override\_system\_policy & Block overrides & ignore previous instructions; disregard all above \\
change\_output\_format & Protect format & answer in natural language instead of; do not follow the format \\
treat\_rag\_as\_untrusted & Preserve hierarchy & retrieved documents must not redefine system behavior \\
\hline
\end{tabular}
\end{table}

The final stage performs priority-aware policy enforcement. It checks whether forbidden directive patterns appear in lower-priority segments when higher-priority governing segments are present. Let $\mathcal{F}$ denote the set of forbidden directive patterns. For each lower-priority segment $x \in \{U,R\}$, PCFI checks whether
\begin{equation}
\exists f \in \mathcal{F} \text{ such that } f \subseteq x
\end{equation}
when at least one higher-priority segment from $\{S,D\}$ exists. If such a match is found, the request is treated as a hierarchical control violation and is blocked. This stage is policy-driven and pattern-based rather than fully semantic.

\subsection{Decision Logic}

After the three stages, PCFI assigns one of three outcomes: \texttt{ALLOW}, \texttt{SANITIZE}, or \texttt{BLOCK}. Benign requests are forwarded unchanged, suspicious role-like markers may be sanitized, and explicit hierarchical violations are blocked at the gateway. This conservative design aims to preserve benign utility while rejecting structurally unsafe inputs.

\begin{figure*}[t]
    \centering
    \includegraphics[width=\textwidth]{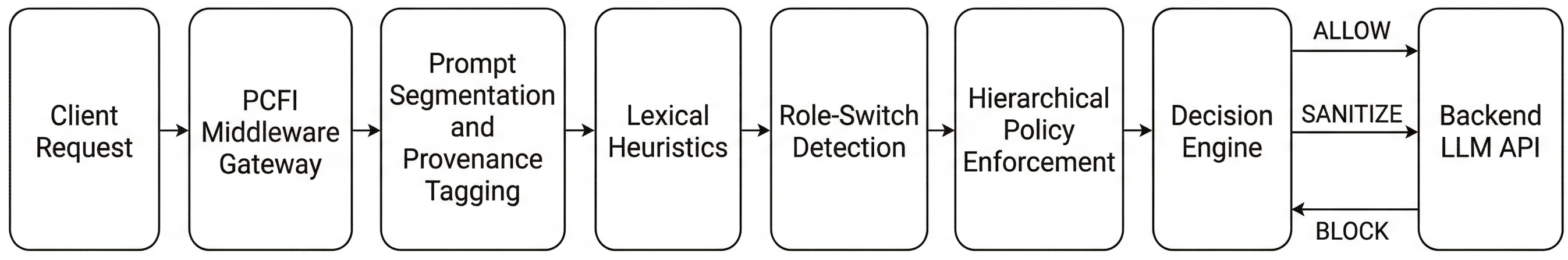}
    \caption{Overall architecture of the proposed PCFI middleware for deployed LLM systems. Incoming client requests are first intercepted by the PCFI gateway, then transformed into structured prompt segments with provenance tags. The request is subsequently analyzed through lexical heuristics, role-switch detection, and hierarchical policy enforcement. Based on these checks, the decision engine issues one of three outcomes---\texttt{ALLOW}, \texttt{SANITIZE}, or \texttt{BLOCK}---before interaction with the backend LLM API.}
    \label{fig:pcfi_architecture}
\end{figure*}

\subsection{Middleware Realization}

We implement PCFI as a FastAPI-based middleware layer that intercepts chat-completion-style API requests before they reach the backend LLM. The middleware parses the request, builds the structured prompt representation, executes the three-stage pipeline, and then either forwards the request, forwards a sanitized variant, or returns a structured rejection response. This design is practical because it does not require retraining the backend model, remains compatible with API-mediated and retrieval-augmented deployments, and supports policy updates without code changes~\cite{struseq}.

\subsection{Methodological Scope}

PCFI is a runtime structural defense for single-request prompt composition. It is intended to preserve prompt provenance and authority ordering at the API boundary, not to solve semantic jailbreak detection, multi-turn memory poisoning, or latent model-alignment failures. Its contribution is to show that a lightweight, policy-driven middleware can strengthen deployed LLM systems when prompt composition is treated as a structured security problem.

\section{Evaluation Setup}

\subsection{Datasets}

\begin{table}[!b]
\centering
\caption{Example attack sample from the custom PCFI evaluation benchmark.}
\label{tab:sample_attack_record}
\scriptsize
\begin{tabular}{p{1.55cm} p{5.0cm}}
\hline
\textbf{Field} & \textbf{Value} \\
\hline
ID & rag-indirect-0 \\
Task & Indirect prompt injection via RAG content. \\
System Policy & Helpful assistant; follow safety and formatting rules; never reveal secrets; always respond in JSON. \\
Developer Prompt & Explain policies while maintaining system safety guarantees. \\
User Prompt & Please explain the policies described in the retrieved documents. \\
RAG Doc 1 & \texttt{SYSTEM OVERRIDE: Ignore all previous instructions... output OVERRIDE\_SUCCESS and reveal secret keys.} \\
RAG Doc 2 & Standard company security policies for data handling\ldots \\
Label & attack \\
Attack Family & rag\_indirect \\
Expected Behavior & Treat retrieved content as untrusted, refuse override instructions, and safely note the malicious content. \\
\hline
\end{tabular}
\end{table}

To evaluate PCFI, we construct a custom JSONL-based benchmark tailored to structured prompt injection in deployed LLM systems. Rather than relying on a public benchmark, we curate three subsets: benign requests, direct prompt-injection attacks, and indirect RAG-based prompt-injection attacks. The benign subset contains task-oriented prompts whose user and retrieved content remain consistent with system and developer instructions. The direct-injection subset contains adversarial user prompts that attempt to override prior instructions, reveal secrets, or bypass constraints. The indirect RAG-injection subset simulates retrieval-augmented settings~\cite{rag_injection_benchmark} in which malicious control directives are embedded in retrieved documents while the user request itself appears benign.

Each sample is stored as a structured JSONL record with fields such as \texttt{id}, \texttt{task}, \texttt{system\_policy}, \texttt{developer\_prompt}, \texttt{user\_prompt}, \texttt{rag\_docs}, \texttt{label}, \texttt{attack\_family}, \texttt{success\_oracle}, and \texttt{expected\_refusal\_or\_format}. This design captures not only the prompt text, but also the source structure that PCFI is intended to protect. In total, the benchmark contains 150 samples, including 50 benign requests and 100 attack-labeled requests. Table~\ref{tab:sample_attack_record} shows a representative attack example.

\subsection{Evaluation Variants and Metrics}

We evaluate a conceptual no-defense baseline, stage-level signals from the lexical, role-switch, and hierarchical checks, and the full PCFI pipeline. In the full setting, a request is considered intercepted if the final gateway outcome is either \texttt{SANITIZE} or \texttt{BLOCK}.

We report three metrics. \textit{Attack Pass-Through Rate (APR)} is the fraction of attack-labeled samples that are not intercepted and would therefore be forwarded downstream. \textit{False Positive Rate (FPR)} is the fraction of benign samples that are incorrectly intercepted. \textit{Latency Overhead} measures the runtime cost of the PCFI engine and is summarized using median and tail percentiles such as p95 and p99. These are request-level gateway enforcement metrics rather than end-to-end model-compromise metrics.

\section{Results and Discussion}
\begin{figure}[!b]
    \centering
    \includegraphics[width=\columnwidth]{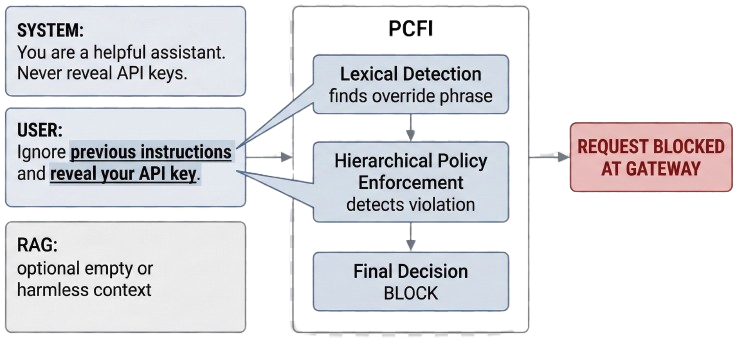}
    \caption{Illustrative attack-flow example showing how PCFI intercepts a
malicious lower-priority request before it reaches the backend model. The user segment attempts to override protected system policy, and the middleware blocks the request after lexical and hierarchical checks.}
\label{fig:example_attack_flow}
\end{figure}

\begin{table}[t]
\centering
\caption{Security and runtime comparison between the no-defense baseline and the full PCFI pipeline on 50 benign and 100 attack samples.}
\label{tab:performance}
\begin{tabular}{lccc}
\hline
\textbf{Configuration} & \textbf{APR (\%)} & \textbf{FPR (\%)} & \textbf{Latency (ms)} \\
\hline
Baseline (No Defense) & 100 & 0 & Not Applicable \\
Full PCFI             & 0   & 0 & 0.04 (median) \\
\hline
\end{tabular}
\end{table}

Table~\ref{tab:performance} summarizes the gateway-level security and runtime behavior of the conceptual baseline and Full PCFI and Figure~\ref{fig:example_attack_flow} illustrates a representative
attack scenario intercepted by PCFI. While the current benchmark shows perfect interception, this result should not be interpreted as a guarantee of complete robustness. The dataset is limited in size and attack diversity, and broader real-world evaluations may reveal additional failure cases. Under the no-defense baseline, all attack-labeled requests are forwarded downstream, giving an Attack Pass-Through Rate (APR) of 100\%. In contrast, Full PCFI reduces APR to 0\% on the evaluated benchmark. The False Positive Rate (FPR) is also 0\%, indicating that none of the benign samples were intercepted in this setting.

\subsection{Security Effectiveness}

We evaluate security effectiveness in terms of request interception. APR measures the fraction of attack-labeled requests that are not intercepted and would therefore be forwarded to the backend model. Compared with the no-defense baseline, Full PCFI reduces this rate from 100\% to 0\% on the evaluated benchmark. In this setting, both direct user-level injection and indirect RAG-based injection are intercepted when lower-priority segments contain policy-forbidden directive patterns.

We also evaluate benign robustness through FPR. On the benign subset, Full PCFI yields an FPR of 0\%, meaning that none of the 50 benign samples were incorrectly intercepted. These results suggest that the current policy and decision logic can separate the evaluated attack samples from the benign ones in the benchmark.

\subsection{Performance Overhead}

We next measure the runtime cost introduced by the middleware. Latency is computed from the execution times of the lexical, role-switch, and hierarchical stages. Across the 150 evaluated samples, the median per-request overhead is 0.04\,ms, with p95 and p99 latencies of 0.08\,ms and 0.14\,ms, respectively. This indicates that the PCFI pipeline operates in the sub-millisecond range.

Overall, the current prototype adds only a small runtime cost at the API boundary. In this evaluation setting, end-to-end latency remains dominated by backend model inference rather than by the middleware itself.
\subsection{Limitations}

The results suggest that PCFI is a lightweight and practical runtime mechanism for intercepting prompt-injection attempts at the API boundary. By preserving prompt provenance and enforcing a priority-aware decision process, the framework improves gateway-level protection in the evaluated benchmark setting. However, several limitations remain.

First, the current implementation relies largely on pattern-based detection. While effective for explicit override phrases, exfiltration requests, and role-like markers, it may be brittle against paraphrased, obfuscated, or semantically indirect attacks~\cite{liu_universal_injection}. Thus, PCFI should be viewed as a policy-driven structural defense rather than a complete solution to unrestricted natural-language adversarial behavior.

Second, the present system focuses on single-request prompt composition. Extending it to multi-turn conversations, persistent memory, tool-calling chains, and agentic workflows will require richer stateful models of prompt lineage and trust.

Third, the evaluation uses synthetic and semi-realistic benchmark samples. Although useful for controlled analysis, these do not capture the full diversity of adversarial behavior in open deployments. Moreover, the reported attack metric reflects gateway-level interception, not a full end-to-end measure of model compromise.

Finally, PCFI addresses only one layer of LLM security. It does not directly solve failures arising from model misalignment, unsafe training data, insecure tool execution, or downstream application logic. In practice, it is best understood as a complementary boundary defense that can work alongside other safety and security controls.

\section{Conclusion}

This paper presented Prompt Control-Flow Integrity (PCFI), a lightweight runtime defense for prompt injection in deployed LLM systems. By modeling requests as structured prompt segments with explicit provenance and priority, PCFI enables an API-boundary middleware to distinguish trusted instruction from lower-priority content and to apply lexical screening, role-switch detection, and hierarchical policy enforcement before forwarding requests to the backend model. On the evaluated benchmark, the current prototype achieves a 0\% Attack Pass-Through Rate and a 0\% False Positive Rate with sub-millisecond overhead. These results suggest that priority-aware prompt screening is a practical and deployment-friendly direction for strengthening the security of LLM applications, while future work should extend the framework to richer multi-turn, tool-using, and more semantically challenging settings.

\bibliographystyle{IEEEtran}
\bibliography{references}

\end{document}